\documentclass[doublecol]{epl2}

\usepackage{epsfig,amsmath,graphicx,graphics,float}

\title{Sum rules, dipole oscillation and spin polarizability of a
spin-orbit coupled quantum gas}

\shorttitle{Sum rules, dipole oscillation and spin polarizability of
a spin-orbit coupled quantum gas}

\author{Yun Li \and Giovanni Italo Martone \and Sandro Stringari}
\shortauthor{Yun Li \etal}

\institute{Dipartimento di Fisica, Universit\`{a} di Trento and
INO-CNR BEC Center, I-38123 Povo, Italy, EU}


\abstract{ Using a sum rule approach we investigate the dipole
oscillation of a spin-orbit coupled Bose-Einstein condensate
confined in a harmonic trap. The crucial role played by the spin
polarizability of the gas is pointed out. We show that the lowest
dipole frequency exhibits a characteristic jump at the transition
between the stripe and spin-polarized phase. Near the second order
transition between the spin-polarized and the single minimum phase
the lowest frequency is vanishingly small for large condensates,
reflecting the divergent behavior of the spin polarizability. We
compare our results with recent experimental measurements as well as
with the predictions of effective mass approximation.}

\begin{document}

\maketitle

Synthetic gauge fields represent a rapidly developing direction of
research in ultra cold atomic physics both from the experimental
\cite{Lin2009_1,Lin2009_2,Lin2011_1,Lin2011_2,Aidelsburger2011,
Hauke2012,LeBlanc2012,Chen2012v1,Wang2012,Zwierlein2012} and
theoretical perspective \cite{Dalibard2010,Liu2009,Stanescu2008,
Wang2010,Wu2011,Sinha2011,Hu2012,Ho2011,Li2012,Ozawa2012_1,
Ozawa2012_2,Vyasanakere2011_1,Vyasanakere2011_2,Gong2011,Hu2011,
Yu2011}. They give rise to the occurrence of new quantum phases
exhibiting unique magnetic features. In the presence of spin-orbit
coupling also the collective oscillations exhibit challenging
features which have been already the object of theoretical studies
\cite{Bijl2011,Zhang2012,Zhang2011,Ramachandhran2012,ChenZhu2012},
as well as first experimental measurements \cite{Chen2012v1}. In
particular the experiment of \cite{Chen2012v1} has shown that the
center-of-mass oscillation of a harmonically trapped Bose-Einstein
condensate can be deeply affected by the coupling with the spin
degree of freedom. In the same paper the problem was investigated
theoretically using a variational approach within time dependent
Gross-Pitaevskii theory.

The purpose of the present work is to study the behavior of the
center-of-mass oscillation employing a sum rule approach. This
method is known to point out the role of conservation laws and to
reduce the calculation of the dynamical properties of the many-body
system to the knowledge of a few key parameters relative to the
ground state. In the case of the center-of-mass oscillation we will
show that, due to the spin-orbit coupling, a key role is played by
the spin polarizability of the gas which is responsible for
important deviations of the dipole frequency from the harmonic
oscillator value.

In this work we consider a spin $1/2$ interacting gas characterized
by the Hamiltonian (for simplicity we set $\hbar=m=1$)
\begin{equation}
H = \sum_i h_0(i) + \sum_{\alpha,\,\beta} \frac{1}{2} \int
d^3\mathbf{r} \, g_{\alpha\beta}\, n_\alpha(\mathbf{r})
n_\beta(\mathbf{r}) \label{H}
\end{equation}
where $i=1,\, \cdots, \, N$ is the particle index, while $\alpha$
and $\beta$ are the spin indices ($\uparrow,\downarrow \,=\pm$)
characterizing the two spin states. The single-particle Hamiltonian
$h_0$ is defined by
\begin{equation}
h_0= \frac{1}{2}\left[\left(p_x-k_0 \sigma_z\right)^2+
p_\perp^2\right] + \frac{\Omega}{2} \sigma_x + \frac{\delta}{2}
\sigma_z +V_{\textrm{ext}}({\bf r}) \label{h0}
\end{equation}
and is characterized by equal contributions of Rashba
\cite{Rashba1984} and Dresselhaus \cite{Dresselhaus1955} spin-orbit
couplings and a uniform magnetic field in the $(x,\,z)$-plane, with
$\Omega$ the Raman coupling constant accounting for the transition
between the two spin states, $k_0$ the momentum transfer of the two
Raman lasers, $\delta$ the energy difference between the two
single-particle spin states, and $V_{\textrm{ext}} =(\omega_x^2 x^2
+ \omega_y^2y^2+\omega_z^2z^2)/2$ the external trapping potential,
thereby chosen of harmonic type. The spin up and down density
operators entering Eq.~(\ref{H}) are defined by $n_\pm({\bf
r})=(1/2)\sum_i\left(1\pm\sigma_{z,i}\right) \delta({\bf r}-{\bf
r}_i)$ while $g_{\alpha\beta}=4\pi a_{\alpha\beta}$ are the relevant
coupling constants in the different spin channels, with
$a_{\alpha\beta}$ the corresponding $s$-wave scattering lengths.
Finally $\sigma_k$, with $k=x,y,z$, are the usual $2\times 2$ Pauli
matrices. The Hamiltonian (\ref{H}), (\ref{h0}) has been already
implemented experimentally \cite{Lin2011_2,Chen2012v1}. It has been
recently employed to describe a variety of non trivial quantum
phases in Bose-Einstein condensates \cite{Ho2011,Li2012}.

In the absence of spin-orbit coupling ($k_0=0$) the center-of-mass
oscillation of the trapped gas along the $x$-direction is exactly
excited by the operator $X = \sum_i x_i$ and corresponds to the
usual sloshing mode with frequency $\omega_x$. In order to
investigate the effect of spin-orbit coupling it is useful to
introduce the following moments of the excitation strengths of the
operator $F$ at zero temperature \footnote{At finite temperature,
the moments $m_k$ should include the proper Boltzmann factors
\cite{book}}:
\begin{equation}
m_k(F)= \sum_n (E_n-E_0)^k\,|\langle0|F|n\rangle|^2 \label{mk}
\end{equation}
where $|0\rangle$ and $|n\rangle$ are, respectively, the ground
state and the excited states of $H$ with $E_n-E_0$ the corresponding
excitation energies, while $|\langle0|F|n\rangle|^2$ is the
$F$-strength relative to the state $|n\rangle$. Some relevant sum
rules can be easily calculated employing the closure relation and
the commutation rules involving the Hamiltonian of the system. For
example the most famous energy-weighted moment for the dipole
operator $F=X$ takes the model independent value (also called
$f$-sum rule) $m_1(X)=(1/2) \langle0|\left[X,\left[H,X\right]\,
\right]|0 \rangle = N/2$ where $N$ is the total number of atoms.
Notice that this sum rule is not affected by the spin terms in the
Hamiltonian, despite the fact that the commutator of $H$ with $X$
explicitly depends on the spin-orbit coupling:
\begin{equation}
\left[H,X\right]= -i\left(P_x-k_0\Sigma_z\right) \, .
\label{commutator}
\end{equation}
Here $P_x= \sum_i p_{x,i}$ is the total momentum of the gas along
$x$ and $\Sigma_z = \sum_i\sigma_{z,i}$ is the total spin operator
along $z$. Equation (\ref{commutator}) actually reflects the fact
that the equation of continuity (and hence in our case the dynamic
behavior of the center-of-mass coordinate) is deeply influenced by
the coupling with the spin variable. The above commutation rule will
be later used to exploit the importance of such a coupling in the
evaluation of the $m_{-3}(X)$ sum rule.

Another important sum rule is the inverse energy-weighted sum rule,
also called dipole polarizability. In the presence of harmonic
trapping this sum rule can be calculated in a straightforward way
using the commutation relation $\left[H,P_x\right]=i\omega^2_x X$
and the closure relation. One then finds the result $m_{-1}(X)=
N/(2\omega^2_x)$, thereby showing that both the energy-weighted and
the inverse energy-weighted sum rules relative to the dipole
operator $X$ are insensitive to the presence of the spin terms in
the single-particle Hamiltonian (\ref{h0}), as well as to the
two-body interaction. This does not mean, however, that the dipole
dynamic structure factor is not affected by the spin-orbit coupling.
This effect is accounted for by another sum rule, particularly
sensitive to the low energy region of the excitation spectrum: the
inverse cubic energy-weighted sum rule for which we find the exact
result
\begin{equation}
m_{-3}(X)= \frac{N}{2 \omega^4_x}\left(1+k^2_0\chi\right)
\label{m-3}
\end{equation}
where we have introduced the spin polarizability per particle
$\chi=2m_{-1}(\Sigma_z)/N$ relative to the $z$-th direction of the
spin operator. In order to derive result (\ref{m-3}) we have used
the recurrence relation $m_{-3}(X) = m_{-1}(P_x)/\omega^4_x$
following from the commutation relation for $\left[H,P_x\right]$, as
well as the most relevant commutation rule (\ref{commutator}) for
$\left[H,X\right]$. It is worth mentioning that the above results
for the sum rules $m_1(X)$, $m_{-1}(X)$ and $m_{-3}(X)$ hold exactly
for the Hamiltonian (\ref{H}), including the interaction terms.
Their validity is not restricted to the mean-field approximation and
is ensured for both Bose and Fermi statistics, at zero as well as at
finite temperature. In particular the sum rule $m_{-3}(X)$, being
sensitive to the magnetic susceptibility, is expected to exhibit a
non trivial temperature dependence across the BEC transition.

Equation (\ref{m-3}) exploits the crucial role played by the
spin-orbit coupling proportional to $k_0$. The effect is
particularly important when the spin polarizability takes a large
value. A large increase of $\chi$ is associated with the occurrence
of a dipole soft mode as can be inferred by taking the ratio between
the inverse and cubic inverse energy-weighted sum rules $m_{-1}(X)$
and $m_{-3}(X)$, yielding the rigorous upper bound $\omega_x/
\sqrt{1+ k^2_0\chi}$ to the lowest dipole excitation energy. The
effect of the coupling between the center-of-mass oscillation and
the spin degree of freedom is further revealed by making the ansatz
$F = P_x + \eta k_0\Sigma_z$ for the optimized operator exciting the
dipole oscillation and minimizing the collective frequency fixed by
the ratio
\begin{equation}
\omega^2= \frac{m_1(F)}{m_{-1}(F)} = \frac{- 2\eta^2 k_0^2 \Omega
\langle\sigma_x\rangle + \omega_x^2}{1 + (1+\eta)^2 k_0^2 \chi}
\label{omega}
\end{equation}
with respect to variations of the real parameter $\eta$. In deriving
the last equality we have explicitly used the sum rule result
$m_{1}(\Sigma_z)= (1/2)\langle0| \left[\Sigma_z, \left[H,
\Sigma_z\right]\right]|0\rangle = -N \Omega \langle\sigma_x \rangle$
for the energy-weighted moment relative to the spin excitation
operator $\Sigma_z$. The choice $\eta=0$ in Eq.~(\ref{omega})
corresponds to the estimate $m_{-1}(X)/m_{-3}(X)$. In the opposite
$\eta \gg 1$ limit Eq.~(\ref{omega}) instead coincides with
$m_1(\Sigma_z)/m_{-1}(\Sigma_z)$. Result (\ref{omega}), based on the
linear response formalism of sum rules, provides a rigorous upper
bound to the dipole frequency in the regime of small amplitude
oscillations.

Let us now discuss the behavior of the physical quantities $\chi$
and $\langle\sigma_x\rangle$ entering the expression for the dipole
frequency. Their actual behavior depends on the quantum phase
characterizing the ground state of the many-body system. These
phases were investigated in details in \cite{Li2012} for a spin
$1/2$ interacting Bose-Einstein condensate employing the ansatz
\begin{equation}
\psi=\sqrt{\frac{N}{V}}\left[C_+ \begin{pmatrix} \cos\theta_+
\\ -\sin\theta_+\end{pmatrix} e^{ik_+x} + C_-
\begin{pmatrix}\sin \theta_- \\ -\cos\theta_-\end{pmatrix}
e^{ik_-x}\right] \label{spinor}
\end{equation}
for the order parameter. The variational calculation, applied to a
uniform configuration, yields the conditions $k_\pm=\pm k_1$ and
$\theta_\pm =\theta= \arccos(k_1/ k_0)/2$ and predicts the
occurrence of three quantum phases, depending on the value of the
relevant parameters $k_0$, $\Omega$ and $g_{\alpha\beta}$ (here and
in the following we consider a spin symmetric Hamiltonian with
$g_{\uparrow\uparrow}=g_{\downarrow\downarrow}=g$ and $\delta=0$).
These are the stripe phase (also called phase I) where Bose-Einstein
condensation takes place in a combination of plane waves with
opposite momenta $\pm k_1$, provided $g>g_{\uparrow\downarrow}$, the
spin-polarized phase II and the single minimum ($k_1 =0$) phase III.
A natural generalization of the variational wave function can be
also used to calculate the dynamic properties of uniform matter, by
solving the Bogoliubov equations and determining the relevant
elementary excitations of the system, the momentum distribution and
the quantum depletion of the condensate \cite{trento2}.

\begin{figure}
\centering
\includegraphics[scale=0.5]{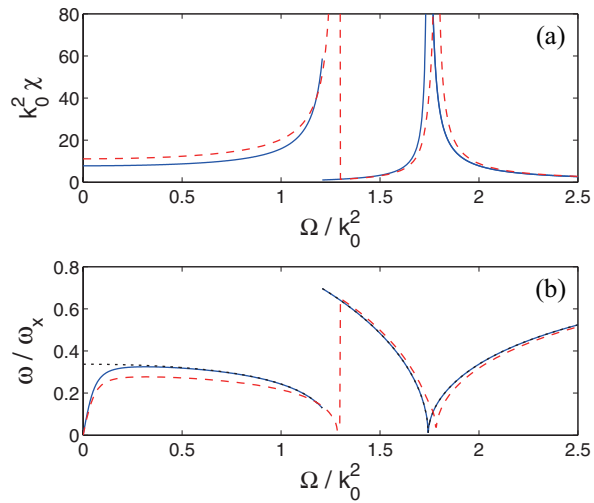}
\caption{(Color online) (a) Spin polarizability $\chi$ as a function
of $\Omega$ calculated in uniform matter (blue solid lines) and in
the harmonic trap (red dashed lines). (b) The corresponding lowest
mode frequency $\omega$ with $\langle\sigma_x \rangle$ and $\chi$
calculated in uniform matter (blue solid lines) and in the harmonic
trap (red dashed lines). The black dotted lines correspond to the
prediction $\omega_x/\sqrt{1+k_0^2 \chi}$ (see text). The
parameters: $k_0^2 = 2\pi\times 320\,$Hz, $\omega_x = 2\pi\times
20\, $ Hz, density in the center of the trap $n \simeq 2.6\times
10^{13}\, $cm$^{-3}$, scattering length $a_{\uparrow\uparrow}
=a_{\downarrow \downarrow} =100\, a_B$, $a_{\uparrow\downarrow}=
60\, a_B$, where $a_B$ is Bohr radius, corresponding to $G_1/k_0^2
\simeq 0.257$ and $G_2/k_0^2 \simeq 0.064$. } \label{fig:chi_omega}
\end{figure}

The static spin polarizability of the spinor Bose-Einstein
condensate at zero temperature can be directly calculated by
expanding the values of $k_\pm$ and $\theta_\pm$ around the
equilibrium values and minimizing the energy of the system in the
presence of an external static field proportional to $\sigma_z$. For
densities smaller than the critical value $n^{(c)}=k^2_0/(2\gamma
g)$, with $\gamma= (g- g_{\uparrow \downarrow})/(g+g_{\uparrow
\downarrow})$, all the three phases discussed above are available
\cite{Li2012}. In the stripe phase, holding for small Raman
frequency $\Omega$, one finds the result
\begin{equation}
\chi_{\textrm{I}}(\Omega) = \frac{\Omega^2 - 4 k_0^4}{\left(G_1 + 2
G_2\right) \Omega^2 - 8 G_2 k_0^4}\,, \label{chiI}
\end{equation}
where we have defined $G_1 = n \left(g + g_{\uparrow \downarrow}
\right)/4$ and $G_2 = n \left(g - g_{\uparrow\downarrow}\right)/4$
and, for simplicity, we have considered the weak coupling limit
characterized by the condition $G_1, \, G_2 \ll k_0^2$. The spin
polarizability $\chi_{\textrm{I}}$ diverges as one approaches the
critical frequency $\Omega^{\textrm{(I-II)}}=2 k_0^2 \sqrt{2
\gamma/(1+2\gamma)}$ providing the transition to the spin-polarized
phase. However Eq.~(\ref{chiI}) is valid only in the low-density
limit and inclusion of higher-order terms makes the value of $\chi$
finite at the transition. In the spin-polarized phase II the spin
polarizability is given by
\begin{equation}
\chi_{\textrm{II}}(\Omega) = \frac{\Omega^2}{\left(k_0^2
-2G_2\right) \left[4 \left(k_0^2-2G_2\right)^2 - \Omega^2\right]}
\,. \label{chiII}
\end{equation}
It takes a larger and larger value as one approaches the transition
point to the zero momentum phase III occurring at the frequency
$\Omega^{\textrm{(II-III)}}= 2( k_0^2-2G_2)$. In the zero momentum
phase III the spin polarizability takes instead the value
\begin{equation}
\chi_{\textrm{III}}(\Omega)=\frac{2}{\Omega - 2\left(k_0^2-2G_2
\right)} \label{chiIII}
\end{equation}
exhibiting a divergent behavior when one approaches the transition
point from above and vanishing in the limit of large $\Omega$.
Results (\ref{chiII}) and (\ref{chiIII}), whose validity is not
limited to the weak coupling regime, explicitly reveal the second
order nature of the phase transition, the values of $\chi$ differing
by a factor $2$ when one approaches the transition from above or
below. They also show that the relevant interaction parameter in
both phases is the spin density dependent parameter $G_2$. The
behavior of the spin polarizability as a function of the Raman
coupling $\Omega$ is shown in Fig.\ref{fig:chi_omega} (a) for a
choice of parameters which emphasizes the deviations with respect to
the weak coupling result. In particular the transition point between
the phases II and III is predicted to be located at a value $\sim
15\, \%$ smaller than the value $2 k_0^2$ (the weak coupling
result). In the same figure we report the value of the spin
polarizability calculated within Gross-Pitaevskii theory in the
presence of 1D harmonic trapping \cite{Li2012}. The comparison with
the uniform matter prediction, corresponding to the value of the
density calculated in the center of the trap, is rather good
reflecting the usefulness of the uniform matter calculation of
$\chi$.

The behavior of the average transverse spin polarization $\langle
\sigma_x \rangle$ was calculated in \cite{Li2012}. In the stripe and
in the spin-polarized phase the uniform matter calculation yields,
respectively, $\langle \sigma_x \rangle = \left(\Omega/2\right)/
\left( k^2_0+G_1\right)$ and $\langle \sigma_x \rangle = \left(
\Omega/2 \right)/\left(k^2_0-2 G_2\right)$, while in the single
minimum phase one has $\langle \sigma_x \rangle = -1$.

We are now in a position to estimate the dipole frequency by
minimizing Eq.~(\ref{omega}) with respect to the parameter $\eta$.
The variational procedure actually provides two solutions. The upper
solution is physically meaningful only for very small values of
$\Omega$ where it approaches the frequency $\omega_x$ of the
center-of-mass sloshing mode. For higher $\Omega$ the upper solution
takes large values and the coupling with other modes, not accounted
for by our ansatz for the excitation operator $F$, becomes
important. The results for the lowest dipole solution are reported
in Fig.\ref{fig:chi_omega} (b). They reveal the important deviations
from the oscillation frequency $\omega_x$ caused by the spin-orbit
and Raman couplings for all values of $\Omega$. In the same figure
we also show the prediction $\omega_x/ \sqrt{1+ k^2_0\chi}$ for the
dipole frequency, obtained by setting $\eta=0$ in (\ref{omega}).
This turns out to be an excellent estimate except for very small
values of the Raman coupling. Actually, in the $\Omega \ll \omega_x$
limit the value of $\eta$ minimizing Eq.~(\ref{omega}) is no longer
small and the mode turns out to be a pure spin oscillation, its
frequency vanishing linearly with $\Omega$. In this limit the mode
does not exhibit any significant coupling with the center-of-mass
oscillation. At the transition between the phases I and II the
lowest dipole frequency exhibits a sudden jump and then starts
decreasing for larger values of $\Omega$. In the thermodynamic
limit, it vanishes at the transition between the phases II and III
as a consequence of the divergent behavior of $\chi$ and, above the
transition, it increases to reach asymptotically the oscillator
value $\omega_x$ at large $\Omega$.

\begin{figure}
\centering
\includegraphics[scale=0.5]{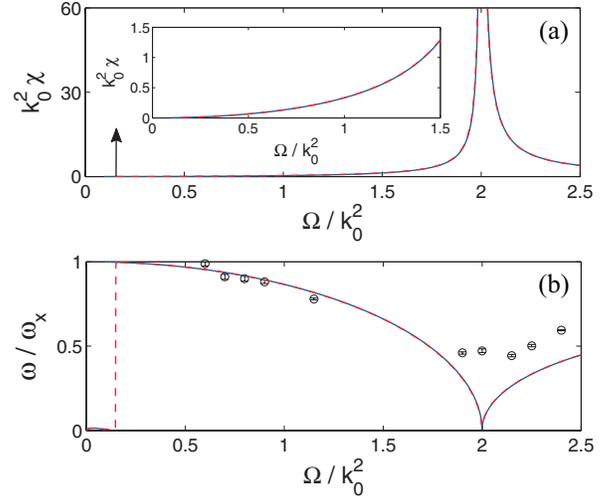}
\caption{(Color online) (a) Spin polarizability $\chi$ and (b)
lowest mode frequency $\omega$ as functions of $\Omega$ (see
Fig.\ref{fig:chi_omega} for details) with different choice of
parameters: $k_0^2 = 2\pi\times 4.42\,$kHz, $\omega_x = 2\pi\times
45\, $ Hz, density in the center of the trap $n \simeq 1.37\times
10^{14}\, $cm$^{-3}$, scattering length $a_{\uparrow\uparrow}
=a_{\downarrow \downarrow} =101.20\, a_B$, $a_{\uparrow\downarrow}=
100.99\, a_B$, where $a_B$ is Bohr radius, and the atomic mass of
$^{87}$Rb. The black circles are the experimental data of
\cite{Chen2012v1}. The black arrow in (a) indicates the transition
between the phase I and II.} \label{fig:chi_omega_exp}
\end{figure}

In Fig.\ref{fig:chi_omega_exp} we show the spin polarizability and
the frequency of the lowest dipole mode calculated in the
experimental conditions of \cite{Chen2012v1}. They correspond to a
very small value of $\gamma$ ($\sim 10^{-3}$), thereby causing the
compression of the stripe phase into a narrow region at small values
of $\Omega$. Furthermore these conditions correspond to a very small
value of $G_2$. As a consequence, in the wide range of Raman
coupling $\Omega \gg \omega_x$ Eqs.~(\ref{omega}), (\ref{chiII}) and
(\ref{chiIII}) yield the useful results
\begin{equation}
\omega^2_{\textrm{II}}= \omega_x^2 \left(1- \Omega^2 /4k_0^4\right),
\; \; \; \omega^2_{\textrm{III}}= \omega_x^2 \left( 1-2k_0^2/
\Omega\right) \label{simpleomega}
\end{equation}
for the dipole frequency in the phases II and III.
Fig.\ref{fig:chi_omega_exp} (a) shows in a clear way the divergent
behavior of the spin polarizability at the transition between the
two phases which is responsible for the quenching of the dipole
frequency. The GP simulations are practically indistinguishable from
the calculations of $\chi$ in the uniform matter (and hence of
$\omega$) based on uniform matter ingredients. The comparison with
the experimental data for the dipole frequencies, as shown in
Fig.\ref{fig:chi_omega_exp} (b), is good far from the transition
point at $\Omega=2 k_0^2$, while near the transition nonlinear
effects play a major role as discussed in \cite{Chen2012v1} (see
also discussion below).

The combined spin-orbit nature of the lowest dipole mode is also
nicely revealed by the relative amplitudes of the oscillating values
of the center-of-mass position ($A_X$), momentum ($A_P$) and spin
polarization ($A_\Sigma$). These amplitudes can be calculated in the
present approach by writing the many body oscillating wave function
as $|\psi(t)\rangle =e^{i\alpha(t)F}e^{\beta(t)D} |0\rangle$ where
$F=P_x+ \eta k_0\Sigma_z$ is the excitation operator and $D$ is
defined by the commutation relation $\left[H,D\right]=F$, while
$\alpha$ and $\beta$ are parameters whose time dependence can be
obtained through a variational Lagrange procedure. The time
dependence of the relevant quantities $\langle X \rangle$, $\langle
P_X \rangle$ and $\langle \Sigma_z \rangle$ is then easily
calculated by expanding the wave function up to first order in
$\alpha$ and $\beta$. For $\Omega \ll \omega_x$, as discussed above,
the lowest frequency mode is mainly a spin oscillation (large
$\eta$) and one finds that the center-of-mass position is basically
at rest ($A_X\sim 0$), while $A_P/k_0 \sim A_\Sigma$. For larger
values of $\Omega$ the lowest frequency is instead well approximated
by the choice $\eta=0$, and the relationships between the spin,
center-of-mass and momentum amplitudes take the useful form
$A_\Sigma =A_X k_0 \omega_x \chi/\sqrt{1 +k^2_0 \chi}$ and $A_P/k_0=
A_\Sigma \left(1+ k^2_0 \chi \right)/\left(k^2_0\chi\right)$. The
connection between the momentum and spin amplitudes has been already
pointed out in \cite{Chen2012v1}. The above relationships show that,
near the transition between the phases II and III, where the spin
polarizability diverges, the amplitude $A_\Sigma$ of the spin
oscillation can easily take large values, thereby emphasizing the
role of nonlinear effects. These effects are likely at the origin of
the finite values of the dipole frequencies observed at the
transition (see Fig. 2 and discussion in \cite{Chen2012v1}). It is
finally worth mentioning that, despite its strong spin nature, the
lowest frequency mode exhausts almost completely the dipole
polarizability sum rule $m_{-1}(X)$, except in the $\Omega \ll
\omega_x$ region. As a consequence it can be easily excited by
displacing the trapping potential.

In the last part of the work we show that a useful insight on the
lowest dipole oscillation can be also obtained by investigating the
dynamic behavior employing directly the lower branch
\begin{equation}
\epsilon(\mathbf{p}) = \frac{1}{2}\left(p_x^2+p_\perp^2 +k_0^2
\right) -\frac{1}{2} \sqrt{4 p_x^2 k_0^2+\Omega^2}\label{epsilon}
\end{equation}
of the excitation spectrum of uniform matter. The new single
particle Hamiltonian can be used to solve the classical
Hamilton-Jacobi equations in both the linear and nonlinear regimes
\cite{Chen2012v1}. It can be also used to calculate the ratio $
m_1(X)/m_{-1}(X)$ between the energy-weighted and inverse
energy-weighted sum rules relative to the dipole operator $X$. While
the result $N/(2\omega^2_x)$ for the dipole polarizability continues
to hold, the energy-weighted moment is deeply affected by the new
Hamiltonian reflecting the projection procedure accounted for by
Eq.~(\ref{epsilon}). By carrying out explicitly the double
commutator one finds the result $\omega^2 = \omega_x^2 \langle
\partial^2_{p_x} \epsilon(\mathbf{p})\rangle$. By replacing the
value of $p_x$ with the value where $\epsilon(\mathbf{p})$ is
minimum, i.e., where Bose-Einstein condensation takes place in
uniform matter (effective mass approximation), one recovers the
results (\ref{simpleomega}) predicted by the sum rule approach in
the weak coupling limit in the phases II and III respectively. A
better estimate of the frequency at the transition point
$\Omega=2k^2_0$ can be obtained by evaluating the integral $ \int
d\mathbf{p} \, n(\mathbf{p})\,\partial^2_{p_x} \epsilon(\mathbf{p})$
with $n(\mathbf{p})$ given by the momentum distribution of the
Gross-Pitaevskii solution in the presence of the harmonic trap. With
the choice of parameters of Fig.\ref{fig:chi_omega_exp} we find
$\omega_{\min}\simeq 0.03\;\omega_x$. The use of lowest branch
dispersion (\ref{epsilon}) turns out to be instead unsuitable to
investigate the lowest dipole frequency in the stripe phase.

In conclusion we have developed  a sum rule description of the
dipole oscillation which explicitly reveals the crucial role played
by the spin polarizability in the presence of spin-orbit coupling.
The method can be also  applied to Fermi gases as well as
generalized to study the coupling between the dipole and other
collective oscillations recently pointed out in \cite{ChenZhu2012}.

Note added after submission: the divergent behavior of the magnetic
susceptibility predicted in the present work has been recently
observed experimentally in \cite{Chen2012v2} through the analysis of
the relative spin and momentum amplitudes.

\acknowledgments

Stimulating discussions with Lev Pitaevskii, Hui Zhai and Shuai Chen
are acknowledged. This work has been supported by ERC through the
QGBE grant.

\end{document}